\documentclass[copyright,creativecommons]{eptcs}
\providecommand{\event}{ICE 2012}

\usepackage{epshare}

\title{Shared Contract-Obedient Endpoints}
\def\titlerunning{Shared Contract-Obedient Endpoints}

\author{Étienne Lozes\institute{Universit\"at Kassel, Germany}
  \and
  Jules Villard\institute{University College London, UK}}
\def\authorrunning{É. Lozes \& J. Villard}

\begin{document}

\maketitle

\begin{abstract}
  Most of the existing verification techniques for message-passing
  programs suppose either that channel endpoints are used in a linear
  fashion, where at most one thread may send or receive from an
  endpoint at any given time, or that endpoints may be used
  arbitrarily by any number of threads. The former approach usually
  forbids the sharing of channels while the latter limits what is
  provable about programs.  In this paper we propose a midpoint
  between these techniques by extending a proof system based on
  separation logic to allow sharing of endpoints. We identify two
  independent mechanisms for supporting sharing: an extension of
  fractional shares to endpoints, and a new technique based on what we
  call reflexive ownership transfer.  We demonstrate on a number of
  examples that a linear treatment of sharing is possible.
\end{abstract}

\section*{Introduction}

One way to tackle the formal verification problem for message-passing
programs is to prove that they implement protocols expressible in a
higher-level formalism and easier to reason about. Two main
approaches coexist: session types on the one
hand~\cite{st-orig}, used to police interactions in programs expressed
either in the $\pi$-calculus~\cite{SessionTypes} or in a
message-passing variant of Java~\cite{SessionJ}, and
\emph{channel contracts} on the other
hand~\cite{cfsm83}, used for instance to describe the protocols in the
Sing\# programming language~\cite{Sing} developed for the Singularity
operating system~\cite{singularity}. High-level protocol
descriptions allow the program verification effort to be split
between checking properties of the protocol itself, such as the
absence of message reception errors (occurring when a message of the
wrong type is received) or of orphan messages (those sent at some
point in the interaction but never received before the channel is
closed), and checking obedience of the program to the protocol.

Most of the existing verification techniques for message-passing
programs suppose that channel endpoints are used in a linear fashion:
no two threads may ever try to send or receive simultaneously on the
same channel endpoint. Indeed, checking the conformance to a given
protocol locally for each thread or process often gives unsound
results if this linearity condition is broken. This limitation
prevents some programs from being proven. For instance, such linear
channel communications are known to enforce a form of
determinism~\cite{Sassone} that excludes standard synchronisation
primitives such as locks and semaphores.  Almost none of the
verification techniques supporting shared channels is able to give
meaningful protocols to them.

This work extends a previous approach based on a marriage of
separation logic and channel contracts~\cite{VLC-aplas09}, which
forbade any active sharing of endpoints, to deal with endpoint sharing
and give meaningful protocols to their interactions. As in previous
work, we consider a simple imperative language that features explicit
memory manipulation and primitives for creating and destroying
bi-directional, asynchronous channels made of two endpoints, and for
sending and receiving messages on individual endpoints. Each message
is composed of a user-defined label (describing the kind of payload
supposed to be transferred) and zero or more values. Crucially, values
may be pointers to memory cells or even to other endpoints, and thus
sending a message may create sharing of resources and foster data
races.  In the absence of sharing, our previous work was able to prove
obedience to channel contracts and absence of data races, and from
that to deduce the absence of reception errors and of orphan
messages. In this setting, transferring a message that is attached to
a resource is reflected in the logic by transferring ownership of that
resource (the message's \emph{footprint} in memory) to the recipient
of the message.

We propose an extension to the previous proof system that deals with
endpoint sharing and achieves two goals: on the one hand, checking
that the exchange of messages on the channels used by programs obeys
protocols defined by \emph{channel contracts}; 
on the other hand, ensuring the
absence of data races. Our notion of data race freedom allows shared
read access to memory cells, and, as often as possible, shared access
to endpoints for sending and receiving.  The latter significantly
complicates the analysis of communications, and in particular one of
the primary goals of the proof system: to check whether channels obey
their declared contracts.

In order to guarantee that shared endpoints obey their contracts, we
extend the previous proof system~\cite{VLC-aplas09} in two ways. The
first extension is based on fractional
shares~\cite{Boyland:SAS03,Bornat:POPL05}, a widespread refinement of
the all-or-nothing ownership in separation logic.  Permissions allow
communications on endpoints that are partially owned, with some
restrictions as we shall see.  Sharing contract-obedient endpoints
with permissions is indeed problematic as a consistent view of the
contract state of the endpoint should be maintained amongst all
sharers.  A correct permission-based sharing thus turns out to be
extremely close in spirit to unrestricted types by Giunti and
Vasconcelos~\cite{GiuntiV10} if slightly more flexible (see the
related works section).  Permission-based sharing still excludes a lot
of interesting protocols.  Consider for instance the following code
featuring a seller and several buyers running in parallel. The seller
bargains the price of its product using endpoint $e$ while several
buyers concurrently access the peer $f$ of $e$:

\begin{center}
\begin{tabular}{@{}c@{\hspace{1em}}|@{\hspace{1em}}c@{\hspace{1em}}|@{\hspace{1em}}c@{}}
\begin{lstlisting}
seller(e) {
 local price = 0;
 while (!good(price)) {
   send(product_descr,e);
   price = receive(offer,e);
 }
}
\end{lstlisting}
&
\begin{lstlisting}
buyer(f) {
 local x;
 receive(product_descr,f);
 x = think_about_it();
 send(offer,f,x);
}

\end{lstlisting}&
\begin{lstlisting}
main {
 (e,f) = open();
 seller(e) ||
  buyer(f) ||
  ... ||
  buyer(f);
}
\end{lstlisting}
\end{tabular}
\end{center}

Here \li{product_descr} and \li{offer} are \emph{labels} identifying
the kind of message being transferred.  A natural contract for the
communication channel $(e,f)$ is $C =
!\li{product_descr}.?\li{offer}.C$, where $!$ is used for sending and
$?$ for receiving from the point of view of the first endpoint, the
other endpoint following the dual contract where $!$ and $?$ are
swapped (notice that contracts are only concerned with the labels and
directions of the messages being exchanged and not their
values). However, the above program is not provable using fractional
permissions or unrestricted types and the reason, as we shall see, is
that this contract contains two distinct states.

The second extension of our proof system precisely addresses this
problem.  The crucial step towards proving this seller/buyers example
is to consider that none of the buyers makes any assumption on the contract
state of $f$ at the beginning of the protocol.  In particular, they
attempt to receive a product description on $f$ without relying on $f$
being in a contract state that ensures that such a message will be
available next.  The crucial argument that validates this reasoning is
that the received message justifies \emph{a posteriori} the buyer
waiting for a \li{product_descr} message.  In other words, the
knowledge (or ownership) of the contract state of $f$ is established
in the post state of receiving $f$ but not in the pre state: the
footprint of this message contains the endpoint that it is received on.

We fully formalise the two extensions of our proof system in the
companion technical report~\cite{lsv-11-23}. In particular, we give a
formal proof of soundness that ensures several safety properties for
proved programs. Due to space limitations we have chosen to keep the
presentation rather informal, focusing on example programs, and we
invite the interested reader to refer to our technical report for more
details on the formalisation.

\paragraph{Outline}
In the first section we recall our previous model of Sing\# and define
some of the main problems of interest for verification.  In the second
section we give a tutorial introduction to the kind of annotations
that are used in our proof system.  In the third section we illustrate
the principle of permission-based sharing of endpoints on a few
examples.  In the fourth section we introduce the principle of linear
sharing with several examples of non-confluent concurrent programs. In
the fifth section we give some highlights of the formalisation of our
proof system. We conclude with related works.

\section{Copyless Message Passing}

\subsection{Toy Programming Language}

We consider a programming language which offers some basic support for
procedures, variables, and threads. Threads share a global memory and
may allocate, deallocate, and manipulate heap objects allocated in
this shared memory. Moreover, threads may synchronise thanks to
message exchanges across heap-allocated communication
channels. Communications are asynchronous and channels can be thought
of as pairs of perfect FIFO buffers. More precisely, channels always
consist of exactly two endpoints and each endpoint can be used to send
to and receive from the other one.  A send instruction
\li{send(m,e,$v_1$,$\dots$,$v_n$)} triggers the emission of a message
tagged with a label \li m emitted on an endpoint \li e with $n$ values
$v_1$, \dots, $v_n$ (the arity of the message is fixed by the tag \li
m).  A receive instruction \li{(x$_1$,$\dots$,x$_n$) = receive(m,e)}
is symmetric and specifies which message tag is expected. This
communication model is close to the one of Sing\#~\cite{Sing}.

\begin{exple}\label{ex:sendendpoint}
The two threads below collaboratively close the channel they use to
communicate. The variables \li e and \li f are global.
\begin{center}
\begin{tabular}{@{}c@{\hspace{.5cm}}|@{\hspace{.5cm}}c@{\hspace{.5cm}}|@{\hspace{.5cm}}c@{}}
\begin{lstlisting}
main(){
 (e,f) = open();
 put() || get();
}

\end{lstlisting}&
\begin{lstlisting}
put() {
 send(endpoint,e,e);
}


\end{lstlisting}
&
\begin{lstlisting}
get() {
 local x;
 x = receive(endpoint,f); 
 close(x,f);
}
\end{lstlisting}
\end{tabular}
\end{center}
The main function allocates the two endpoints \li e, \li f of the
channel with \li{open} and gives one to each of the two threads
\li{put} and \li{get} that it launches in parallel. It
then joins both threads.  The \li{put} thread sends a message
tagged as ``\li{endpoint}'' to the \li{get} thread.
Observe that \li e is both the subject and the object of this message:
endpoints are heap-allocated and the address of an endpoint is a valid
value for a message.  Upon reception the \li{get} thread holds
both endpoints of the channel and can thus safely close the channel
with \li{close}.
\end{exple}

Our toy programming language also features primitives for heap
manipulation. For simplicity, we will assume primitive heap cells with
only two fields. We use the syntax \li{x = new()}, \li{dispose(x)},
\li{x.$i\ $=$\ \dots$}, and \li{y = x.$i$} for allocation,
deallocation, destructive update, and look-up ($i\in\{0,1\}$ denotes
the field selector). Although endpoints are also heap cells, they are
of a different type and we only consider programs that are well-typed
in that respect (the proof system will actually verify that this is
the case).  Message values can contain the memory address of an
endpoint or a cell and thus both kind of heap objects can be passed by
message exchanges. The grammar below summarises some of the key
aspects of the syntax of our programming language.
$$
\begin{array}{rcl}
  c &::=& \li{skip} ||| \li{x =$\ e$} ||| \null\\
  &&\li{x = new()}
  ||| \li{x.0 = y} ||| \li{x.1 = y} ||| 
\li{x = y.0} |||\li{x = y.1} |||
  \li{dispose(x)} ||| \null\\
  &&\li{(x,y) = open($C$)} ||| \li{close(x,y)}
   ||| \li{send(m,x,x$_1$,$\dots$,x$_n$)} |||
   \li{(x$_1$,$\ldots$,x$_n$) = receive(m,x)} ||| \null \\
   && \li{switch \{ $\ \ldots\ $ case $\ \vec{\tx{x}_i}\ $ = receive(m$_i$,e$_i$): $\ p_i$ $\ \ldots\ $ \}}
   ||| \ldots\\\\
  p &::=& c ||| \li{$p$;$\;p$} ||| \li{$p$||$p$} ||| \li{while ($b$) $\ p$} |||
  \li{if ($b$) then $\ p\ $ else $\ p$} ||| \li{local x;$\;p$} ||| \ldots
\end{array}
$$

\begin{exple}\label{ex:ex2}
In the code below the \li{put} thread non-deterministically chooses
either to send a cell or to keep it. On the other side, the receiver
(\li{get}) needs to be ready to receive any of the two messages
\li{cell} and \li{no_cell}; this is achieved by the
\li{switch}/\li{case} construct. Variables not declared as local are
global.
\begin{center}
\begin{tabular}{c@{\hspace{.5cm}}|@{\hspace{.5cm}}c@{\hspace{.5cm}}|@{\hspace{.5cm}}c}
\begin{lstlisting}
main(){
 (e,f) = open();
 x = new();
 put() || get();
}






\end{lstlisting}&
\begin{lstlisting}
put() {
 local f;
 if ($\ldots$) {
  send(cell,e,x); 
 } else {
  send(nocell,e); 
  dispose(x);
 }
 f = receive(clos,e);
 close(e,f);
}
\end{lstlisting}&
\begin{lstlisting}
get() {
 switch {
 case y = receive(cell,f):
  dispose(y);
 case receive(nocell,f):
  skip;
 }
 send(clos,f,f);
}


\end{lstlisting}
\end{tabular}
\end{center}
\end{exple}

\subsection{Pitfalls of Copyless Message Passing}
\label{sec:problems}
Programs may suffer from several runtime errors.  First of all,
\emph{memory violations} may occur when a dangling pointer is
dereferenced, modified, or disposed. Moreover, two threads may cause a
\emph{data race} if they simultaneously try to access a variable or a
memory location and if at least one of the accesses is a write access.
Synchronisations via message passing can be used to prevent such races
but might yield \emph{deadlocks}: in this paper, a program is
considered to be in a deadlock configuration if all of its threads are
blocked on receptions.

In addition to these rather familiar problems, it is important to
mention other erroneous behaviours one may wish to avoid. The first
one is \emph{memory leaks} (situations where it becomes impossible to
fully deallocate the heap), which one may want to detect or forbid
altogether, either because the underlying programming language is not
garbage collected or because this information may matter for the
garbage collector (the latter holds for Sing\#).  Similarly, sent
messages may never be received and stay in a channel forever; closing
channels when they contain such \emph{orphan messages} is most of the
time considered a communication error. Instead, one should make sure
that both queues of a channel are empty before closing it.  Lastly, a
thread may enter a \li{switch}/\li{case} construct in a situation
where the buffer of one of the endpoints it scans starts with a
message whose tag is not listed as a possible case. We strive to
ensure the absence of such \emph{unspecified receptions}, though they
might be resolved differently in other communication models (for
instance the program may block or look for a message with one of the
listed tag deeper in the queue, \emph{\`a la} Erlang).

\begin{exple}\label{ex:safety-problems}
The program
$
P_0 == \li{send(m$_1$,e) || switch \{case receive(m$_2$,f): skip\}}
$
is both deadlocking and triggering an unspecified
reception. Similarly, replacing \li{get()} by \li{skip}
in Example~\ref{ex:sendendpoint} would cause a memory leak. 
\end{exple}

Note that all of these problems are distinct and incomparable.
In particular,
\begin{itemize}
\item orphan messages are not a special case of memory leaks: it
  may be the case that lost messages carried no ownership
  (\textit{e.g.} the message \li{nocell} above);
\item unspecified receptions are not a special case of deadlocks: we
  will distinguish the program $P_1 ==
  \li{send(m$_1$,e) || receive(m$_2$,f)}$ from the program $P_0$
  above: $P_1$ is only deadlocking.\footnote{Provided no other thread
    accesses \li{f} concurrently. Note that one may wish to
    authorise a thread to wait for a message in a buffer provided that
    other threads will pick all the other ones first.  In order to
    support such behaviours we distinguish reception from scanning
    (also called peeking) and consider that unspecified receptions
    might occur only on scanning, not on reception.}  Conversely, the
  program \li{$P_0\ $ || \{receive(m$_1$,f); send(m$_2$,e);\}} is
  not deadlocking.  However, it has an unspecified reception at the
  time \li m$_1$ is sent.
\end{itemize}

\section{Proving Copyless Message Passing}

\subsection{Separation Logic}
Our aim is to design a proof system for Hoare triples based on
separation logic that will prevent all of the aforementioned errors
except deadlocks\footnote{The reason why we omit deadlocks is not that
  we find them uninteresting, but rather that there is no simple way
  to treat them without introducing new notions; see Leino et
  al.~\cite{Leino:ESOP2010} or session types~\cite{SessionTypes} for
  analyses of message-passing programs dealing with deadlocks.}, as
was achieved previously in a context without endpoint
sharing~\cite{VLC-aplas09}.  Hoare triples are understood in terms of
partial correctness and, for this reason, non-terminating and
deadlocking programs generally have a proof. A triple
$\metahoare{A}{p}{B}$ is informally understood as ``starting $p$ from
any state satisfying $A$ will not result in a memory fault, an
unspecified reception or an orphan message and, if $p$ terminates,
then the final state satisfies $B$''.

Being based on separation logic, our proof system will ensure that
every proved program follows some locality principles.  As a first
approximation (and before we introduce permissions), these locality
principles can be summarised as follows.\footnote{The ownership
  hypothesis and separation property also apply to global
  variables. Since this is cumbersome to formalise we leave that
  aspect implicit in this paper.}
\begin{description}
\item[Ownership hypothesis] Each thread owns a subpart of the heap: it
  can only read and update that part of the heap. The part of the heap
  owned by a thread may evolve during the execution.
\item[Separation property] At any point in the execution each allocated cell
  is owned by either exactly one thread or by a message stored in
  a queue.
\end{description}

These principles can be illustrated on the program of
Example~\ref{ex:sendendpoint}: observe that at every point in time,
\li{put} and \li{get} own disjoint parts of the heap.
Initially, \li{put} owns the endpoint \li{e} while
\li{get} owns the endpoint \li{f}.  During execution,
the ownership of \li{e} is transferred from \li{put} to
the message \li{endpoint} and ultimately to \li{get},
which justifies its disposal by \li{get}.

State assertions make this reasoning more formal.  They convey
information on the part of the heap that is owned by the thread at a
given point in time.  The base predicates of state assertions describe
the ownership of a single cell or endpoint.

The example code on the left-hand side of Figure~\ref{fig:ex-annot1}
features the annotations that formalise the explanations given above.
Separation logic formulas appear in square brackets. In the
precondition of \li{main}, $\emp$ indicates that nothing is
known to be allocated (or, alternatively, that nothing in the heap is
owned at this program point).  After the \li{(e,f) = open()}
instruction, two peer endpoints have been allocated and their
addresses have been stored in variables \li{e} and
\li{f} respectively.  The predicate $e \emapsto f$ denotes the
ownership of the endpoint $e$ and provides the information that its
peer is $f$.  In the second assertion, the separating conjunction $*$
is used to add together the two disjoint pieces of owned heap
consisting each of one endpoint.  The proof of the parallel
composition distributes the pieces of heap currently owned to each of
the sub-threads and merges them back at the end of the parallel
composition.

The main limitation of this treatment of parallelism is that it
excludes programs based on the multiple-readers/single writer
principle.  One solution is to assign to every owned piece of heap a
permission $@p\in (0,1]$. The permission $1$ is the ``total'' or
  ``write'' permission while a permission $@p<1$ is a ``partial'' or
  ``read'' permission.  To illustrate this idea, consider the example
  code on the right-hand side of
  Figure~\ref{fig:ex-annot1}. Initially, the program allocates a new
  cell $x$ for which it may assume a write permission. It results in a
  state where the ownership of the cell $x$ with permission $1$ can be
  assumed, which is denoted by $x|->(-,-)$.  It is then allowed to
  update the content of the cell, resulting in
  $x|->(-,7)$.\footnote{We write $x|->(v_1,v_2)$ for a cell $x$ whose
    first and second fields contain the values $v_1$ and $v_2$. We
    freely use a wildcard ``$-$'' when a value is existentially
    quantified. Note that we use different arrows for regular memory
    cells ($|->$) and endpoints ($\emapsto$).}  At the level of the
  parallel composition, we split the write permission to distribute it
  on each side, resulting in two read permissions which we choose to
  be $0.5$ each (uneven splittings would also have been possible), for
  which we write $x|->_{0.5} (-,7)$.\footnote{We explicitly specify
    the subscript permission only when it is different from $1$.}  On
  each side, cell updates are now forbidden (note that all sharers may
  in return assume that the value of the second field does not change)
  but shared reads are allowed.  After the parallel step, all
  permissions are collected back and a write permission for the cell
  is again granted which allows us to dispose the cell.

\begin{figure}
\centering
\begin{tabular}{c@{\hspace{.5cm}}|@{\hspace{.5cm}}c}
\begin{lstlisting}
main() [$\emp$] {
  (e,f) = open();
  [$e\emapsto f * f\emapsto e$]
  [$e\emapsto f$]$\;$ || [$f\emapsto e$]
  put()  || get();
  [$\emp$]     || [$\emp$]
} [$\emp$]




\end{lstlisting}&
\begin{lstlisting}
multi_readers() [$\emp$] {
  x = new(); 
  [$x |-> (-,-)$]
  x.1 = 7;
  [$x |-> (-,7)$]
  [$x |->_{0.5} (-,7)$]$\;$  || [$x |->_{0.5} (-,7)$]
  z = 1 + x.1 || t = 3 $\times$ x.1
  [$x |->_{0.5} (-,7)$]$\;$  || [$x |->_{0.5} (-,7)$]
  [$x |-> (-,7)$]
  dispose(x);
} [$\emp$]
\end{lstlisting}
\end{tabular}
\caption{Separation logic at work: state splitting and parallel
  composition.}
\label{fig:ex-annot1}
\end{figure}

\begin{figure}
\centering
\begin{tabular}{c@{\hspace{.5cm}}|@{\hspace{.5cm}}c}
\begin{lstlisting}
message endpoint(x) [$x\emapsto f$]

put() [$e\emapsto f$] {
  send(endpoint,e,e);
} [$\emp$]


\end{lstlisting}&
\begin{lstlisting}
get() [$f\emapsto e$] {
  local x;
  x = receive(endpoint,f);
  [$f\emapsto e * x\emapsto f$]
  [$f\emapsto x * x\emapsto f$]
  close(x,f);
} [$\emp$]
\end{lstlisting}
\end{tabular}

\caption{Proof sketches for \li{put} and \li{get}.}
\label{fig:proof-putget}
\end{figure}

To conclude this overview, let us complete the proof of the \li{put}
and \li{get} functions of Example~\ref{ex:sendendpoint}.  The
\emph{footprint} of a message is the piece of heap that is lost when
sending this message and gained when receiving it.\footnote{We use a
  different terminology than previous work~\cite{VLC-aplas09} where
  footprints were called \emph{message invariants}.} The correctness
of the interaction between \li{put} and \li{get} is based on the
assumption that the footprint of the \li{endpoint} message is the
endpoint \li{e}, represented by the formula $e \emapsto f$. More
generally, given a program to prove, we will associate a footprint
description to every user-defined message tag $m$ and assume that
every time $m$ is sent or received the same footprint is
transferred. Figure~\ref{fig:proof-putget} shows the state assertions
at each program point for \li{put} and \li{get}, which constitute a
proof sketch. Note that before closing the channel the proof makes the
deduction step that $f\emapsto e * x\emapsto f |- f\emapsto x *
x\emapsto f$, which is valid because each endpoint is peered to
exactly one other endpoint.

\subsection{Channel Contracts}
\label{sec:contracts}

Separation logic naturally enforces some of the safety conditions we
listed in Section~\ref{sec:problems}, such as data race
freedom. However, in order to detect either an orphan message or an
unspecified reception, proofs must support assumptions about the
behaviour of the environment of a thread. In particular, one should be
able to specify which messages local endpoints \emph{must} be ready to
receive and which messages they \emph{may} send.

Channel contracts are a specialised form of rely/guarantee reasoning
that precisely addresses this problem; they describe the protocol
followed by each endpoint as a finite-state automaton.  A contract $C$
is written from one of the endpoints' point of view, the other
endpoint being assumed to follow the dual contract $\bar C$ where
sends $!$ and receives $?$ have been swapped.

\begin{exple}\label{ex:contracts}
The contracts for the two previous example programs are depicted
below.

{\rm\null\hfill
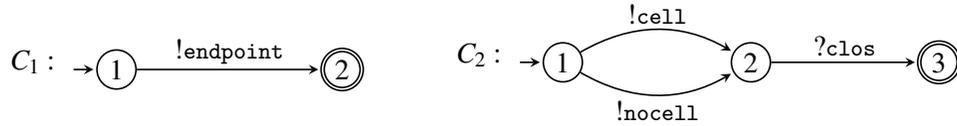
\begin{minipage}[c]{.3\linewidth}
$C_1:$
\begin{tikzpicture}[node distance=3cm]
  \node[state,initial] (1) {1};
  \node[state, accepting, right of=1] (2) {2};

  \path (1) edge [above] node {!\li{endpoint}} (2);
\end{tikzpicture}
\end{minipage}\hfill
\begin{minipage}[c]{.45\linewidth}
$C_2:$
\begin{tikzpicture}[node distance=2.5cm]
  \node[state,initial] (1) {1};
  \node[state,right of=1] (2) {2};
  \node[state, accepting, right of=2] (3) {3};

  \path (1) edge [above, bend left]node {!\li{cell}} (2);
  \path (1) edge [below, bend right] node {!\li{nocell}} (2);
  \path (2) edge [above] node {?\li{clos}} (3);
\end{tikzpicture}
\end{minipage}\hfill\null}

The first contract says that exactly one message \li{endpoint}
will be sent from \li{e} to \li{f}, after which the
channel may be closed. The second contract says that either
\li{cell} or \li{nocell} will be sent and this should be
answered by a \li{clos} message, after which the channel can be
closed.
\end{exple}

Contracts are used to ``type'' endpoints: at any point in time, every
endpoint is decorated with a contract and a contract state. The
contract is fixed for the whole life of the endpoint, from \li{open}
to \li{close}. The contract state evolves along the communications: if
\li{e} is in the contract state $q$ of its contract $C$ and if a
message $m$ is sent over \li{e}, then the contract state of \li{e}
becomes the $!m$ successor of $q$ in $C$. If $q$ does not have such a
successor, then the contract is violated.  Similarly, in case of a
reception of $m$ the contract state evolves to its $?m$ successor. A
\li{switch}/\li{case} construct on a given endpoint should be ready to
handle at least all the receptions indicated by the contract in the
current control state of the endpoint. Peer endpoints \li{e}, \li{f}
allocated with \li{open($C$)} are ruled by contracts $C$ and $\bar C$
respectively and start in the initial contract state of $C$.  Finally,
\li{close(e,f)} is only allowed when both endpoints are peers of each
other and are in the \emph{same} final state of their contract
(contracts may have multiple final states that way).  Note that
contracts without final states describe persistent channels that
cannot be closed.

\begin{exple}\label{ex:formal-ownership}
The proof that the endpoint-transferring program of
Example~\ref{ex:sendendpoint} obeys the contract $C_1$ of the
Example~\ref{ex:contracts} is sketched below. Annotations now feature
the predicate $\speer e C q f$ that denotes the ownership of an
endpoint $e$ ruled by the contract $C$, currently in state $q$ of $C$,
and whose peer is $f$.

\noindent\centering
\begin{tabular}{@{}c|c|c@{}}
\begin{lstlisting}
message endpoint(x) [$\speer x {C_1} 2 {f}$]
put() [$\speer e {C_1} 1 f$] {
 send(endpoint,e,e);
} [$\emp$]

\end{lstlisting}&
\begin{lstlisting}
main() [$\emp$] {
 (e,f) = open($C$); 
 put() || get(); 
} [$\emp$]

\end{lstlisting}&
\begin{lstlisting}
get() [$\cpeer f {C_1} 1 e$] {
 local x;
 x = receive(endpoint,f);
 close(x,f);
} [$\emp$]
\end{lstlisting}
\end{tabular}
\end{exple}

In general, contracts do not guarantee the absence of communication
errors or of orphan messages. Moreover, they are known to be Turing
powerful so deciding most properties of interest is an undecidable
problem~\cite{ws-fm11}. However, sufficient conditions exist that
ensure their correctness: if contracts are deterministic automata with
no mixed choice (all outgoing transitions from a given state are
either all sends or all receives), then receptions errors are
prevented.  If moreover every cyclic path containing a final state
contains at least one send transition and one receive transition, then
there cannot be orphan messages upon channel
closure~\cite{ws-fm11}. All the contracts in this paper satisfy these
restrictions.

\section{Contract Obedience beyond Linearity}

Permissions support scenarios involving sharing among multiple
readers.  Until now we only saw how permissions apply to standard heap
cells, but not endpoints. Let us introduce a predicate $\speersh e
{@p} C q f$ that denotes the ownership of a fraction $@p$ of the
endpoint $e$.  This extension immediately raises a question: what
permission do we require to send, receive, and close?  A conservative
solution which would again prevent any active sharing would be to
require a full permission in all cases.  A rather naive analogy with
heap cells could suggest to ask for a full (write) permission for
\li{send} and \li{close} but only a partial (read) permission for
receptions. Were we concerned with routing messages deterministically
we could conversely allow partial permission for sending but require
full permissions for receiving.  Our solution is more permissive than
both of these: we only require full permissions to close channels.

This choice comes with some restrictions, lest the proof system
becomes unsound. When two threads concurrently send (or receive) the
messages $m_1$ and $m_2$ respectively over a shared endpoint \li{e}
they may change the contract state of \li{e} in different ways and get
inconsistent views of the actual endpoint's state.  Would it be safe
if we were to restrict ourselves to situations where $m_1=m_2$?  The
answer is no and is worth being illustrated by an invalid
proof. Consider a cell-transferring protocol with two producers
sharing an endpoint \li{e} ruled by the contract $C =
\begin{tikzpicture}[baseline]
  \node[circle,draw,initial,inner sep=2pt] (1) {1};
  \node[circle,draw,accepting, right of=1,inner sep=2pt] (2) {2};
  \path (1) edge [above] node {!\li{cell}} (2);
\end{tikzpicture}$.
\begin{center}
\begin{tabular}{@{}c@{\hspace{1em}}|@{\hspace{1em}}c@{\hspace{1em}}|@{\hspace{1em}}c@{}}
\begin{lstlisting}
message cell(x) [$x|-> -$]
put() [$\speersh e {0.5} {C} 1 f$] {
  local x; x = new();
  send(cell,e,x);
} [$\speersh e {0.5} {C} 2 f$]
\end{lstlisting}&
\begin{lstlisting}
main() [$\emp$] {
  (e,f) = open($C$);
  put() || put() || get();
  close(e,f);
} [$\emp$]
\end{lstlisting}&
\begin{lstlisting}
get() [$\cpeer f {C} 1 e$] {
  local y;
  y = receive(cell,f); 
  dispose(y);
} [$\cpeer f {C} 2 e$]
\end{lstlisting}
\end{tabular}
\end{center}

This program should be rejected: there is an orphan message and a
memory leak when the channel is closed. However, the invalid proof
above incorrectly establishes the Hoare triple
$\hoare{\emp}{main()}{\emp}$.  Moreover, both concurrent
\li{put} functions update the contract state of \li{e}
to the same state $2$.

Let us think about the problem differently and consider that
endpoints are typed at run-time (which is the case for Sing\#), or in
other words that the contract state is a field of the heap cell
implementing the endpoint.  Following the same principle as for
standard heap cells, we should only allow updates to the contract
state of an endpoint when it is owned with the full write
permission. In the following we only consider proofs that follow this
principle.  Excluding updates of the contract state of a shared
endpoint may seem to forbid any communication on this endpoint.  There
is however a very particular case where a \li{send} or \li{receive} does not
generate a contract state update: a self-loop on a contract state.

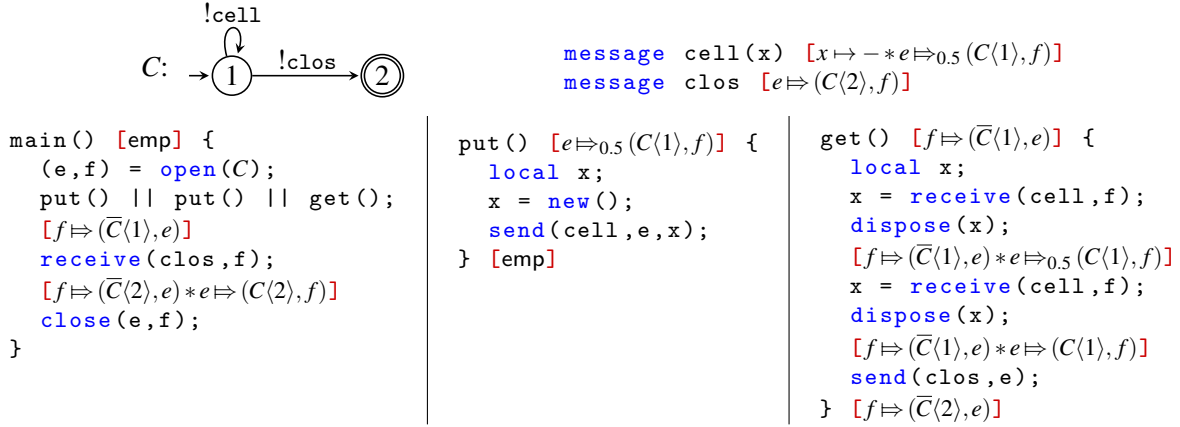
\begin{figure}
\centering
\null\hfill$C$:
\begin{tikzpicture}
  \node[initial,state] (1) {1};
  \node[state,accepting,right of=1] (2) {2};

  \path (1) edge [loop above] node [above] {!\li{cell}} (1);
  \path (1) edge node {!\li{clos}} (2);
\end{tikzpicture}
\hfill
\begin{minipage}{.4\textwidth}
\begin{lstlisting}
message cell(x) [$x|-> - * \speersh e {0.5} C 1 f$]
message clos [$\speer e C 2 f $]
\end{lstlisting}
\end{minipage}\hfill\null

\begin{tabular}{c@{\hspace{1em}}|@{\hspace{1em}}c@{\hspace{1em}}|@{\hspace{1em}}c}
\begin{lstlisting}
main() [$\emp$] {
  (e,f) = open($C$); 
  put() || put() || get();
  [$\cpeer f C 1 e$] 
  receive(clos,f); 
  [$\cpeer f C 2 e * \speer e C 2 f$] 
  close(e,f);
}


\end{lstlisting}&
\begin{lstlisting}
put() [$\speersh e {0.5} C 1 f $] {
  local x; 
  x = new();
  send(cell,e,x);
} [$\emp$]





\end{lstlisting}&
\begin{lstlisting}
get() [$\cpeer f C 1 e$] {
  local x; 
  x = receive(cell,f); 
  dispose(x);
  [$\cpeer f C 1 e * \speersh e {0.5} C 1 f$]
  x = receive(cell,f); 
  dispose(x);
  [$\cpeer f C 1 e * \speer e C 1 f$]
  send(clos,e);
} [$\cpeer f C 2 e$]
\end{lstlisting}
\end{tabular}
\caption{Two producers and one consumer: non-linear sharing.}
\label{fig:twoproducers}
\end{figure}

\begin{exple}\label{ex:twoproducers}
Consider the contract $C$ depicted in Figure~\ref{fig:twoproducers}.
Then a shared endpoint in state $1$ can be used to send a cell message.
The two producers / one consumer program of Figure~\ref{fig:twoproducers}
can thus be proved.
\end{exple}

Note how sending and receiving \li{clos} only occurs with the write
permission on the endpoint, whereas \li{send(cell,e,x)} occurs
with just a read permission over \li{e}. Crucially, the \li{cell}
message silently carries half of the ownership of the endpoint \li{e}.
After receiving a \li{cell} message twice the consumer 
has full ownership of \li{e} and can thus send the
\li{clos} message in a contract-obedient way. 
Note that we cannot avoid the transfer of \li{e}: were the \li{clos}
message sent from \li{f}, the contract state 1 would be a mixed state.

\begin{exple}[Internal Choice]
A slightly more involved example is the following encoding of internal
choice. We let two processes compete to get a \li{token}
message. The one that catches it first can start executing and sends a
\li{notoken} message to the other process, which is in charge
of closing the channel that carried the token.
$$choice(p_1,p_2)==
\li{(e,f) = open($C$); \{ send(token,e) ||$\ p_1'\ $||$\ p_2'\ $\}}
$$
where \li e, \li f do not appear free in $p_1$, $p_2$ and 
$p_i'$ is defined parametrically in the program $p_i$ by
\begin{lstlisting}
switch { case receive(token,f): {send(notoken,e); $p_i$}
         case receive(notoken,f): close(e,f); }
\end{lstlisting}
Let us derive the Hoare triple $\metahoare{A}{choice(p_1,p_2)}{B}$
from $\metahoare{A}{p_1}{B}$ and $\metahoare{A}{p_2}{B}$ for any
formulas $A$ and $B$.  We first give to each process half of the
ownership on the endpoint \li{f}. Then, we consider that the
ownership of the endpoint \li{e} is transferred by the messages
\li{token} and \li{notoken}. Moreover, we set the
message \li{notoken} to transfer half of the ownership of
\li{f}, so that the loser of the internal choice gets the full
ownership on the two endpoints in the end and can close the channel.

{\rm\null\hfill
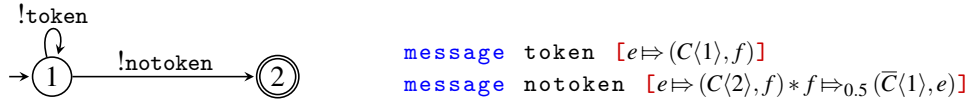
\begin{tikzpicture}[node distance=3cm]
  \node[initial,state] (1) {1};
  \node[state,accepting,right of=1] (2) {2};

  \path (1) edge [loop above] node [above] {!\li{token}} (1);
  \path (1) edge node {!\li{notoken}} (2);
\end{tikzpicture}\hfill
\begin{minipage}{.45\textwidth}
\begin{lstlisting}
message token [$\speer{e}{C}{1}{f}$]
message notoken [$\speer {e}{C}{2}{f} *\cpeersh{f}{0.5}{C}{1}{e}$]
\end{lstlisting}
\end{minipage}
\hfill\null}
\end{exple}

\section{A Linear Form of Sharing}
\label{sec:linear}

Sharing endpoints is very common in scenarios involving one server and
several clients: the server usually owns an endpoint, say \li{ep}, and
the clients share its peer, say \li{ip}. One way to model a client
connection is to assume that the server first sends a \li{welcome}
message. Once caught by one of the clients, a new pair of endpoints is
used for the rest of this connection.  The code in
Figure~\ref{fig:clients_server} models such negotiated connections
among a central server and several clients, letting the clients
allocate the endpoints used for their connection.\footnote{The client
  often uses only few services of the server. For this reason, it
  could be expected that allocating endpoints on the client's side
  achieves better performances, especially if the buffer's size
  drastically shrinks between the client's specific contract and a
  general-purpose service contract implemented by the server.} For
added realism, the code assumes a primitive \li{bind} that takes a
``spinster'' endpoint $\mathit{ip}$ and peers it to a new endpoint. It
could be replaced by an \li{open} primitive in an initial phase before
the clients and the server are spawned. The \li{spawn} construct may
be simulated using $||$ and recursion.

\begin{figure}
\centering
\begin{tabular}{l@{\hspace{1cm}}|@{\hspace{1cm}}l}
\textbf{Clients and Server}&
\textbf{``System calls''}\\
\begin{lstlisting}
server(){
  local e;
  ep = bind(ip);
  listen();
  while ($\ldots$) {
    e = accept();
    spawn serve(e);
  }
  close(ip,ep);
}

client(){
  local f;
  f = connect();
  start_service(f);
}
\end{lstlisting}
&
\begin{lstlisting}
listen(){ send(welcome,ep); }

accept(){
  local e;
  e = receive(connect,ep);
  send(welcome,ep);
  return e;
}

connect(){
  local e,f;
  receive(welcome,ip);
  (e,f) = open($\mathit{Service}$);
  send(connect,ip,e);
  return f;
}
\end{lstlisting}
\end{tabular}

\caption{Negotiating connections between a server and several clients.}
\label{fig:clients_server}
\end{figure}

Observe that the \li{welcome} message carries no endpoint for the
service, but rather grants the right to the client that receives it to
initiate a service. The server thus lets the client allocate the two
endpoints for the service.  One of them is sent back to the server in
the \li{connect} message.  Overall, the connection contract
$\mathit{Service}$ is
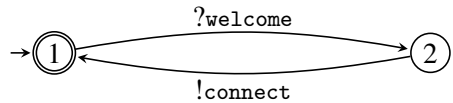
\begin{center}
\begin{tikzpicture}[baseline=(1).base]
    \node[state,initial,accepting] (1) at (0,0) {1}; \node[state] (2) at (5,0) {2};
    \path (1) edge [->,bend left = 10] node[above]{?\li{welcome}} (2);
    \path (2) edge [->,bend left = 10] node[below]{!\li{connect}} (1);
\end{tikzpicture}
\end{center}

Such a contract is problematic for permission-based sharing, as all
transitions have a different source and target state. As we saw in the
previous section, we should not allow communications on partially
owned endpoints in that case.

The problem may be narrowed down to the \li{connect} function, where
the client first receives on $\mathit{ip}$ and then sends on it.
Clearly, the client needs to own the endpoint $\mathit{ip}$ with full
permission in the state before the send, because at this time it has
to change the endpoint's state. Since other clients may be in the
initial state of the \li{connect} function at the same time, not even
a read permission on the endpoint $\mathit{ip}$ can be granted in the
pre state of the \li{connect} function. Between the two states the
message \li{welcome} is received; this seems like a good place to
acquire the ownership of $\mathit{ip}$. But shouldn't the client own
$\mathit{ip}$ before it receives the \li{welcome} message on it? It
indeed needs to update the contract state of $\mathit{ip}$ as well.

The answer is no; let us explain why.  The \li{receive} instruction
needs indeed the ownership of $\mathit{ip}$ but not necessarily in the
pre state. Reasoning backwards gives the explanation. In a correct
proof we would have
$\hoare{\mathit{Pre}}{receive(welcome,ip)}{\mathit{Post}}$ for some
$\mathit{Pre},\mathit{Post}$ formulas.  Since the reception updates
the contract state of $\mathit{ip}$, $\mathit{Post}$ should be of the
form $\cpeer {\mathit{ip}} C 2 {\mathit{ep}} * \ldots$ Let $F$ denote
the footprint of the \li{welcome} message and let
$\mathit{Post}'=\speer {\mathit{ip}} C 1 {\mathit{ep}} * \ldots$
denote the weakest precondition of $\mathit{Post}$ before updating the
contract state. Then we should have that $\mathit{Pre}*F$ entails
$\mathit{Post}'$. As it turns out, nothing more is required and the
ownership of $\speer {\mathit{ip}} C 1 {\mathit{ep}}$ may just as well
be granted by the footprint $F$ of the message. In other words, the
ownership of the endpoint $\mathit{ip}$ can be granted \emph{a
  posteriori} by the reception of the \li{welcome} message over
$\mathit{ip}$.

Let us present the formal proof of the \li{connect} function
using this principle. We will soon present the formal rules supporting
this reasoning and refer to our technical report for a proof of
soundness.
\begin{center}
\begin{tabular}{c@{\hspace{.5cm}}|@{\hspace{.5cm}}c}
\begin{lstlisting}
message welcome [$\peer {\mathit{ip}} C 1 {\mathit{ep}}$]
message connect(e) [$\peer {\mathit{ip}} C 2 {\mathit{ep}} * e\emapsto -$] 





\end{lstlisting}&
\begin{lstlisting}
connect() [$\emp$] {
  local e,f; receive(welcome,ip);
  [$\peer {\mathit{ip}} C 2 {\mathit{ep}}$]
  (e,f) = open($\mathit{Service}$);
  [$\peer {\mathit{ip}} C 2 {\mathit{ep}} * e\emapsto f * f\emapsto e$]
  send(connect,ip,e); return f;
} [$f\emapsto e$]
\end{lstlisting}
\end{tabular}
\end{center}

This way of reasoning might seem unorthodox since it breaks the
(false) intuition that shared channels are not linear.  Perhaps
surprisingly, the above example is indeed a case of sharing channels
in a purely linear fashion.  Before we formalise the way that our
proof system works, let us illustrate the strength of this reasoning
principle on a last example (see also our technical report for
examples with multicast and synchronisation barriers).

\begin{exple}[Locks] 
We develop now an encoding of heap-allocated locks. Our
design choices closely follow the ones of 
Gotsman et al.~\cite{Gotsman}. 

We introduce a macro predicate $\predi{handle}_{\pi}(x)$
($@p\in(0,1]$) that denotes the right to attempt acquiring or
  releasing a lock $x$.  The ownership of the handle is granted with
  full permission when the lock is allocated and it is required with
  full permission as well when the lock is disposed, whereas only a
  read permission is required to either acquire or release the
  lock. We also introduce a macro predicate $\predi{locked}(x)$ that
  holds when the lock is locked by the current thread. It is assumed
  whenever the lock is acquired and lost when the lock is released.
  We consider non-reentrant locks, \textit{i.e.} the formula
  $\predi{locked}(x)*\predi{locked}(x)$ is unsatisfiable (other
  design choices are possible).

The lock intends to protect a piece of heap that satisfies a certain
invariant $I$. When the lock is acquired the ownership of $I$ should
be $*$-conjoined and when it is released $I$ should be consumed. Like
Gotsman et al.~\cite{Gotsman}, we assume that the lock is initially
acquired by the thread that creates it and that it can only be
released by a thread that has acquired the lock. We propose the
following encoding of locking primitives:
\begin{center}
\begin{tabular}{c@{\hspace{.5cm}}|@{\hspace{.5cm}}c}
\begin{lstlisting}
new_lock(x) [$\emp$] {
  x = new(); 
  (x.0,x.1) = open($C$);
} [$\predi{handle}(x)*\predi{locked}(x)$]

dispose_lock(x) [$\predi{handle}(x)*\predi{locked}(x)$] {
  send(stop,x.0); receive(stop,x.1); 
  close(x.0,x.1); dispose(x);
} [$\emp$]
\end{lstlisting}&
\begin{lstlisting}
acquire(x) [$\predi{handle}_{\pi}(x)$] {
  receive(token,x.0);
} [$\predi{handle}_{\pi}(x)*\predi{locked}(x)*I$]

release(x) [$\predi{handle}_{\pi}(x)*\predi{locked}(x)*I$] {
  send(token,x.1);
} [$\predi{handle}_{\pi}(x)$]


\end{lstlisting}
\end{tabular}
\end{center}

The encoding above is based on two messages \li{token} and 
\li{stop}. The first one is used to transfer the ownership of 
the lock from a thread to the next thread that acquires the lock. The second
one triggers the deallocation. 
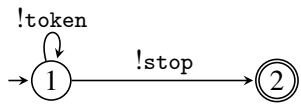
\begin{center}
{\rm\null\hfill
\begin{tikzpicture}[node distance=3cm]
  \node[initial,state] (1) {1};
  \node[state,accepting,right of=1] (2) {2};

  \path (1) edge [loop above] node [above] {!\li{token}} (1);
  \path (1) edge node {!\li{stop}} (2);
\end{tikzpicture}
\hfill
\begin{minipage}{.3\textwidth}
\begin{lstlisting}
message token [$\predi{locked}(x) * I$]
message stop [$\emp$]
\end{lstlisting}
\end{minipage}\hfill\null}
\end{center}
All messages transit through 
the endpoints $x.0$ and $x.1$, which
are thus shared among all client threads. More precisely,
the ownership of the cell $x$ is shared by means of read permissions
but the ownership of the endpoints is shared linearly using the same form
of backward reasoning as in the previous example. The \li{token}
message thus transfers the write ownership of $x.0$ and $x.1$. and 
gives the right to receive the \li{token} message in the
\li{acquire} function even if the endpoint $x.0$ is not owned in the
pre state.  The macro predicates are defined as follows:
$$
\begin{array}{lll}
\lmacr{locked}(x) & == ~~ & \exists X_0,X_1.~x|->_{0.5}(X_0,X_1) * \speer {X_0}{C}{0}{X_1} *
\cpeer{X_1}{C}{0}{X_0} \\
\lmacr{handle}_{\pi}(x) & == & x|->_{\pi/2}(-,-)
\end{array}
$$
\end{exple}

\section{Proof System}

\subsection{Proof Rules}
Our proof system is defined by a set of
inference rules for Hoare triples. The core of the proof system
is composed of the standard proof rules of concurrent separation 
logic~\cite{OHearn04} which are omitted here due to space restrictions. Let us
briefly comment the new rules for copyless message passing.

\paragraph{Channel allocation and deallocation}
The rules for allocation and disposal are rather symmetric: \li{open}
produces two peered endpoints whereas \li{close} consumes them.
\begin{mathpar}
  \infrule{}{q_0=\mathsf{init}(C)}
  {\hoare{\emp}{(e,f) = open($C$)}
    {\speer e C {q_0} f * \cpeer f C {q_0} e}}
  {}
  
  \infrule{}{q\in\mathsf{final}(C)}
  {\hoare{\speer e C q {f} * \cpeer{f} C q e}
    {close(e,f)}{\emp}}{}
\end{mathpar}
Notice that \li{open} allocates two endpoints that are peers and in
the initial state while \li{close} requires two peers that are in a
same final state.  Also note that a full permission is granted and
required in each case.

\paragraph{Send and receive}
The proof rules of \li{send} and \li{receive} require the introduction
of a pseudo-instruction \li{skip$_{e\lambda,f}$}, where $\lambda$
denotes either $!m$ or $?m$. It has the operational meaning of
\li{skip} but also updates the contract state.
\[
  \infrule{}
  {q\stackrel{@l}{\rightarrow}q' \in C
    \\
    q=q'|/\pi=1}
  {\hoare{\speersh e {\pi} {C} q f}
    {skip$_{e\lambda,f}$}
    {\speersh e {\pi} {C} {q'}{f}}}{}
\]
Intuitively, \li{skip$_{e\lambda,f}$} modifies the contract state of
$e$ with respect to the action $@l$ and checks that the transition is
indeed authorised by the contract. Updating a contract state requires
only a partial read permission if the state is left unchanged, and a
total permission otherwise.

Let us now consider the rules for \li{send} and \li{receive}. Let
$I_m(x,y_1,\dots,y_n)$ be the footprint of a message
$m(y_1,\dots,y_n)$ -- the $x$ parameter of the footprint formula
denotes the endpoint that sends the message. Then $I_m(e,\vec{y})$ is
lost upon sending and acquired upon receiving.  Either operation
updates the contract state accordingly using \li{skip$_{e\lambda,f}$}.
\begin{mathpar}
  \infrule{}
  {\hoare{A}{skip$_{e!m,f}$}{B*I_m(e,\vec{x})}}
  {\hoare{A}{send($m$,$e$,$\vec{x}$)}{B}}{}

  \infrule{}
  {\hoare{A*I_m(f,\vec{x})}{skip$_{e?m,f}$}{B}
    \\ \vec x\text{ not free in }A}
  {\hoare{A}{(x$_1$,$\ldots$,x$_n$) = receive($m$,$e$)}{B}}{}  
\end{mathpar}
Note that in the \li{send} case the footprint of the message is
transferred \emph{after} the endpoint state has been updated. Note also 
that the footprint may contain the endpoint itself. Examples of such a reflexive
ownership transfer are the \lstinline{clos} message or the \li{cell}
message in Example~\ref{ex:twoproducers} (the latter with half 
permission).

Conversely, in the \li{receive} case the footprint of the message is
transferred \emph{before} the update of the endpoint's state. The
footprint may again contain the endpoint itself. This possibility
permits linear sharing of endpoints. Examples of such a reflexive
ownership transfer are the \li{welcome} message of the client-server
example, the \li{token} message of the lock example, or the
\li{notoken} message of the encoding of internal choice (the latter
with half permission).

\paragraph{Switch/Case} Finally, two rules address the 
\li{switch}/\li{case} construct: the first rule dispatches switches on
different endpoints into different subproofs. The second rule
addresses \li{switch}/\li{case} on a single endpoint: in that case,
this endpoint must be owned (at least partially) and the switch is
checked to be exhaustive with respect to the possible incoming
messages at the given contract state.
\begin{mathpar}
  \infrule{}
  {\hoare{A}{switch \{ cases1 \}}{B}\\
    \hoare{A}{switch \{ cases2 \}}{B}}
  {\hoare{A}{switch \{ cases1 cases2 \}}{B}}
  {}

  \infrule{}
  {\mathsf{choices}(C,q)\subseteq \{m_1,\dots,m_n\} \\ 
    \hoare{\speersh e {\pi} C {q} {f} * A}
          {$\vec{x_i}\ $= receive($m_i$,$e$):$\;p_i$}
          {B}~\text{for all }i}
  {\hoare{\speersh e {\pi} C {q} {f} * A}
    {switch \{$\ \dots\ \vec{x_i}\ $= receive($m_i$,$e$):$\;p_i\ \dots\ $\}}
    {B}}
  {}
\end{mathpar}

\subsection{Soundness}

Villard defined an operational semantics for our programming language
in his PhD thesis~\cite{Villard-thesis} and showed the soundness of a
previous version of our proof system. The proof easily adapts to the
extended version of the proof system (see our technical
report~\cite{lsv-11-23}).

Many problems complicate the proof of the fact that ``well proved''
programs are safe. By \emph{safe} we mean that all the erroneous
behaviours listed in Section~\ref{sec:problems} are ruled out, except
for deadlocks.  To be well proved, a program does not just need a
proof derivation in our proof system, it also needs well-formed
contracts and well-formed message footprints. Firstly, as mentioned in
Section~\ref{sec:contracts}, contracts should ensure the absence of
communication errors and of orphan messages.  Secondly, we require
that footprints be \emph{precise} formulas (a common restriction in
concurrent separation logics~\cite{csl-viktor-11}). Lastly, all of
these requirements do not suffice to detect all memory leaks.  Sing\#
adopts the following restriction which prevents them: an endpoint can
be sent only when it is in a sending state. This restriction is a
relaxed version of Merro's locality condition for the
$\pi$-calculus~\cite{Merro-phd}. Many of the examples of the paper do
not follow Sing\#'s restriction: the encoding of internal choice and
all the examples of Section~\ref{sec:linear} use the possibility of
sending an endpoint $f$ over its peer $e$.  Generalisations of the
Sing\# condition include the well-foundedness condition of Bono and
Padovani~\cite{Messa:ESOP2011} and Villard's
condition~\cite{Villard-thesis}. In our technical development we use
a condition close to the latter. We refer to the technical report for
the details of our soundness result and proofs.

\section{Related Works}
Giunti and Vasconcelos were the first to consider the problem of
sharing contract obedient endpoints and extended session types for
that purpose by introducing a distinction between linear and
unrestricted channels~\cite{GiuntiV10}.
The linear qualifier applies to endpoints that have to be used
linearly, as in ordinary session types, while the unrestricted one
allows any behaviour on the endpoint.
To retain soundness, unrestrictedly typed channels are limited to
``single state'' protocols and thus this type system can handle
significantly less scenarios than ours. Giunti recently implemented a
type-checker for qualified types~\cite{Giunti2011}.


Turon and Wand~\cite{Ricain} were the first to propose a proof system
for message-passing concurrency that exploited permissions. Their proof system
addresses the (untyped) $\pi$-calculus and introduces some proof rules for
temporal reasoning and refinement.

Leino, Mueller and Smans~\cite{Leino:ESOP2010} proposed a proof system
for reasoning about locks and asynchronous unidirectional
channels. Since they don't have contracts, buffers are assumed
to transfer only one kind of message. Their work
focus instead on preventing deadlocks. 
They implemented their proof system in the Chalice tool.

Bell, Appel and Walker~\cite{Bell:SAS2010} proposed a proof system for
asynchronous unidirectional channels shared by means of permissions.
Their proof system ignores contracts and instead describes the
contents of channels explicitly and relies
on a rule for interleaving
the local histories of two endpoints when they get $*$-conjoined.

Bono, Messa, and Padovani~\cite{Messa:ESOP2011} introduced a type 
system \emph{\`a la} session type for CoreSing\#, a model of Sing\# 
very close in spirit 
to our previous work without sharing~\cite{VLC-aplas09}. They accurately pointed out
that our previous work prevented orphan messages but not
memory leaks, and introduced a well-foundedness condition to fix the problem.  
More recently, they extended their type system with qualified 
types~\cite{typingclmp-journal}.

Hobor and Gherghina~\cite{Hobor:ESOP2011} proposed a proof system for
synchronisation barriers based on separation logic that was helped by
``barrier diagrams'', a contract-like description of the sequences of
synchronisations on the barrier. 
 We propose an encoding of barriers with sharing in our technical 
report~\cite{lsv-11-23}.

Sassone, Rathke and Francalanza~\cite{Sassone} introduced a separation logic
for CCS and stated a confluence
property for the form of linear CCS addressed by their logic.
Our proof system shows that, with little effort, it is
possible to go past confluent programs while still obeying this
kind of discipline. 

Merro showed that the full $\pi$-calculus can be encoded in the local
$\pi$-calculus~\cite{Merro-phd}. On the contrary, the
contract-obedient $\pi$-calculus underlying our model of Sing\# seems
strictly more expressive than the local, contract-obedient one.
However, we did not try to characterise the expressive power of our
unrestrictedly linear $\pi$-calculus.

\bibliographystyle{eptcs}
\bibliography{biblio}

\end{document}